\documentclass[11pt,a4paper]{article}
\usepackage[margin=27mm]{geometry}
\usepackage[T1]{fontenc}
\usepackage[utf8]{inputenc}
\usepackage{lmodern}
\usepackage{amsmath,amssymb,amsthm}
\usepackage{booktabs,array,tabularx,longtable}
\usepackage{enumitem,microtype,needspace}
\usepackage{placeins}
\usepackage{tikz}
\usetikzlibrary{arrows.meta,positioning,calc,fit,backgrounds}
\usepackage{hyperref}
\hypersetup{colorlinks=true,linkcolor=blue!45!black,citecolor=blue!45!black,
  urlcolor=blue!45!black,bookmarksdepth=2,
  pdftitle={Two Rounds Are Already Hard for Loop-Free Updates of Routing Trees},
  pdfauthor={Klaus-Tycho Foerster}}
\definecolor{oldblue}{RGB}{32,76,117}
\definecolor{neworange}{RGB}{170,68,27}
\definecolor{earlyfill}{RGB}{225,240,239}
\definecolor{latefill}{RGB}{250,237,220}
\definecolor{inkgrey}{RGB}{63,72,80}
\tikzset{
  oldedge/.style={-{Stealth[length=1.8mm]},line width=.8pt,oldblue},
  newedge/.style={-{Stealth[length=1.8mm]},line width=.8pt,neworange,dashed},
  routevertex/.style={circle,draw=inkgrey,line width=.65pt,minimum size=7mm,inner sep=1pt,font=\small},
  auxrouter/.style={rounded corners=1.5pt,draw=inkgrey,line width=.65pt,minimum width=10mm,minimum height=7mm,inner sep=2pt,font=\small},
  destination/.style={circle,draw=inkgrey,double,double distance=1pt,minimum size=6mm,inner sep=1pt,font=\small},
  figlabel/.style={font=\small,text=inkgrey},
  figheading/.style={font=\small\bfseries,text=inkgrey},
  restpath/.style={-{Stealth[length=1.6mm]},densely dotted,line width=.8pt,inkgrey}
}

\newtheorem{theorem}{Theorem}[section]
\newtheorem{lemma}[theorem]{Lemma}

\newtheorem{corollary}[theorem]{Corollary}
\theoremstyle{definition}

\theoremstyle{remark}

\setlist{topsep=4pt,itemsep=3pt,parsep=0pt,leftmargin=*}
\makeatletter
\renewcommand{\@openbib@code}{\setlength{\itemsep}{0pt}\setlength{\parsep}{0pt}}
\makeatother

\title{Two Rounds Are Already Hard for Loop-Free Updates of Routing Trees}
\author{Klaus-Tycho Foerster\\{\normalsize TU Dortmund, Germany}}
\date{21 November 2026}

\begin{document}
\maketitle

\begin{abstract}
Updating a routing tree can create loops even when the old and new trees
lead to the same destination. A controller chooses which routers update
in each round, but their updates may finish in any order. Strong loop
freedom therefore requires every partial completion to be acyclic.
We show that deciding whether two rounds suffice is NP-complete, even
when a safe three-round schedule is supplied. The proof is a linear-size
reduction from 3-SAT with one old clause spine and four routers per
variable. Three types of
mixed cycles enforce consistent literal choices and clause satisfaction.
For every nonempty input formula, the constructed instance needs exactly
two rounds if the formula is satisfiable and exactly three otherwise.
\end{abstract}

\section{Introduction}\label{sec:intro}

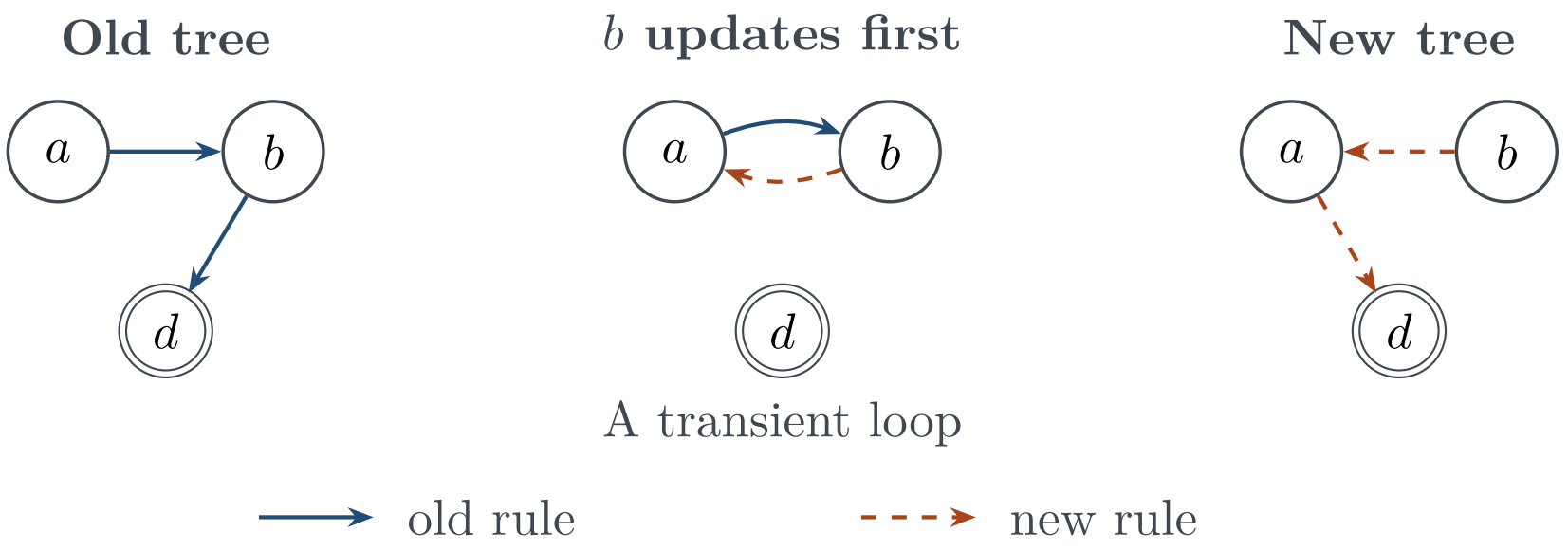
\begin{figure}[htbp]
\centering
\begin{tikzpicture}[x=1cm,y=1cm]
  \begin{scope}[shift={(1.35,0)}]
    \node[figheading] at (0,1.35) {Old tree};
    \node[routevertex] (oa) at (-.75,.55) {$a$};
    \node[routevertex] (ob) at (.75,.55) {$b$};
    \node[destination] (od) at (0,-.7) {$d$};
    \draw[oldedge] (oa)--(ob);
    \draw[oldedge] (ob)--(od);
  \end{scope}
  \begin{scope}[shift={(5.65,0)}]
    \node[figheading] at (0,1.35) {$b$ updates first};
    \node[routevertex] (ma) at (-.75,.55) {$a$};
    \node[routevertex] (mb) at (.75,.55) {$b$};
    \node[destination] (md) at (0,-.7) {$d$};
    \draw[oldedge] (ma) to[bend left=20] (mb);
    \draw[newedge] (mb) to[bend left=20] (ma);
    \node[figlabel] at (0,-1.35) {A transient loop};
  \end{scope}
  \begin{scope}[shift={(9.95,0)}]
    \node[figheading] at (0,1.35) {New tree};
    \node[routevertex] (na) at (-.75,.55) {$a$};
    \node[routevertex] (nb) at (.75,.55) {$b$};
    \node[destination] (nd) at (0,-.7) {$d$};
    \draw[newedge] (nb)--(na);
    \draw[newedge] (na)--(nd);
  \end{scope}
  \draw[oldedge] (2.0,-2.0)--(2.8,-2.0);
  \node[figlabel,anchor=west] at (2.9,-2.0) {old rule};
  \draw[newedge] (6.2,-2.0)--(7.0,-2.0);
  \node[figlabel,anchor=west] at (7.1,-2.0) {new rule};
\end{tikzpicture}
\caption{Both endpoint trees reach $d$, but updating $b$ before $a$
creates the loop $a\to b\to a$. Updating $a$ first and $b$ second is safe.
Solid blue arrows are old rules; dashed orange arrows are new rules.}
\label{fig:motivation}
\end{figure}

We study updates between two routing trees with a common destination.
Each non-destination router replaces its prescribed next-hop rule exactly
once. The controller selects the routers in each round; their updates may
finish in any order~\cite{Foerster19Tutorial}. Although both endpoint trees
are loop-free, intermediate forwarding states can contain loops. A safe
round must therefore remain loop-free after \emph{every} partial completion
(Figure~\ref{fig:motivation}). How many such rounds are necessary?

For source--destination path pairs, two-round feasibility is
polynomial-time solvable, whereas three-round feasibility is NP-complete
under \emph{strong} loop freedom
\cite{FoersterLMS18,LudwigMS15}. Three-round hardness was also known
for routing trees with one destination
\cite[Section~VI]{FoersterLMS18}\cite[Theorem~1]{FoersterW16}.
The corresponding \emph{two}-round problem remained
open~\cite[Section~VI]{FoersterLMS18}. We prove that it is already NP-complete.

\begin{theorem}\label{thm:main}
The following decision problems are NP-complete.
\begin{enumerate}[label=(\alph*)]
\item Given two successor tables on the same finite router set and a
destination, accept if and only if the tables define spanning routing
trees toward that destination and admit a strongly loop-free schedule
using at most two rounds.
\item Given, in addition, a proposed three-round schedule, accept if and
only if the conditions in~(a) hold and the supplied schedule is valid and
strongly loop-free. A valid schedule assigns each non-destination router
to exactly one of rounds~1, 2, and~3; empty rounds are allowed.
\end{enumerate}
NP-hardness in~(b) holds even when the tables define such trees and the
supplied three-round schedule is valid and safe.
\end{theorem}

For every nonempty input formula, the constructed routing-tree instance
needs exactly two rounds if the formula is satisfiable, and exactly three
otherwise. Even with a safe three-round schedule in hand, deciding whether
two rounds suffice is therefore NP-hard. Checking whether all routers can
update in one round requires only an acyclicity test.

The reduction starts from 3-SAT formulas with at least one clause;
this restriction preserves NP-hardness. The formal input reduction
prepends a tautological clause to every well-formed formula, making
even an empty formula nonempty; malformed inputs are mapped to a fixed
unsatisfiable formula.
It uses one old \emph{clause spine} and four routers per variable. Updating a choice router in the first round represents
selecting a literal. Three types of cycles prevent contradictory
selections and require a selected literal in every clause. Conversely,
a satisfying assignment yields a safe two-round schedule. The essential
observation is that a proposed first-round set must pass \emph{two}
acyclicity tests, one for each round.

Section~\ref{sec:model} gives the two-round safety test.
Section~\ref{p:proof} presents the reduction and its correctness proof.
Section~\ref{r:models} derives the approximation consequence and compares
the result with other update models.

\section{Model and Two-Round Test}\label{sec:model}

We work with one destination and two prescribed routing trees, as in
Foerster and Wattenhofer~\cite[Section~III]{FoersterW16}. Strong loop
freedom requires the absence of a cycle anywhere in the forwarding
graph, not only along the path of one designated ingress
\cite[Section~II-A]{FoersterLMS18}.

\paragraph{Forwarding states.}
Let $V$ be a finite set of routers, let $d\in V$ be the destination, and
write $U=V\setminus\{d\}$. The old and new successor functions
$f_0,f_1:U\to V$ are part of the input. For $i\in\{0,1\}$, put
\[
 E_i(Z)=\{(v,f_i(v)):v\in Z\},\qquad E_i=E_i(U).
\]
Both $(V,E_0)$ and $(V,E_1)$ are spanning in-arborescences toward $d$.
Thus every router other than $d$ has one successor, and following
successors from any router eventually reaches $d$. We impose no degree
bound.

A router atomically replaces its old successor with its new one, once.
There are no temporary next hops, extra forwarding rules, or packet
tags. If exactly the routers in $Z\subseteq U$ have completed their
updates, the forwarding graph is
\begin{equation}\label{eq:state}
 F(Z)=\bigl(V,E_1(Z)\cup E_0(U\setminus Z)\bigr).
\end{equation}
Every nonroot router still has exactly one outgoing edge, so an
acyclic state automatically routes every vertex to $d$.

\paragraph{Asynchronous rounds.}
A schedule $(Q_1,\ldots,Q_k)$ partitions $U$ into rounds; empty rounds
are permitted. All routers in one round must finish before the next
round begins. But within round $j$ an arbitrary subset $T\subseteq Q_j$
may already have finished, while the others still use their old rules.
With $P_j=\bigcup_{i<j}Q_i$, the forwarding state is then $F(P_j\cup T)$.
A schedule is \emph{strongly loop-free} if every such state is acyclic,
including the beginning and the end of every round. We consider
instantaneous forwarding states, not packet trajectories across several
states. Assigning a router with identical old and new rules to either
round makes no difference.

\paragraph{A two-round schedule has one choice.}
Let $S\subseteq U$ denote the routers assigned to the first round;
the remaining routers update in the second. Call members of $S$
\emph{early} and the others \emph{late}. To check safety, we consider
all edges that \emph{could} occur during each round:
\begin{equation}\label{eq:two}
\begin{split}
 G_1(S)&=\bigl(V,E_0\cup E_1(S)\bigr),\\
 G_2(S)&=\bigl(V,E_1\cup E_0(U\setminus S)\bigr).
\end{split}
\end{equation}
An early router contributes both its old and new edges to $G_1$, even
though an actual forwarding state uses only one. The same applies to
a late router in $G_2$.

\begin{lemma}[Union criterion]\label{p:union}
A round is strongly loop-free if and only if the union of its possible
forwarding edges is acyclic.
\end{lemma}
\begin{proof}
Every partial forwarding state is a subgraph of the union. Conversely,
if the union has a directed cycle, take a simple such cycle. At each
router on it, the cycle uses only one outgoing edge. We can therefore
choose that router's old or new state, independently of the other
routers in the round, to realize the whole cycle in one partial
completion. Routers outside the round have only one available edge.
\end{proof}

Thus \emph{two rounds are safe exactly when both $G_1(S)$ and $G_2(S)$
are acyclic}. Testing a proposed $S$ takes polynomial time. The same
observation lets us verify a supplied three-round schedule in polynomial
time. Both problems in Theorem~\ref{thm:main} are total decision languages:
malformed tables and endpoints that are not destination trees are rejected.
Part~(b) also rejects invalid or unsafe supplied schedules.

\begin{table}[htbp]
\centering
\caption{Which edges appear in the round-union graphs. ``Both'' means
both edges are available in the union, not simultaneously installed at
one router.}
\label{p:availability}
\begin{tabularx}{\linewidth}{@{}lXX@{}}
\toprule
Router's assigned round & In $G_1(S)$ & In $G_2(S)$ \\
\midrule
Early: round 1 & Old and new & New only \\
Late: round 2 & Old only & Old and new \\
\bottomrule
\end{tabularx}
\end{table}

The two tests pull in opposite directions. Every directed cycle of
$(V,E_0\cup E_1)$ must lose at least one edge in each round. In $G_1$,
this must be an edge of $E_1\setminus E_0$ whose tail is \emph{late};
in $G_2$, it must be an edge of $E_0\setminus E_1$ whose tail is
\emph{early}. Conversely, these two omissions from every cycle suffice
to make both unions acyclic. Edges shared by the old and new tables are
always present and cannot serve either role. The reduction exploits
this tension: one set $S$ must meet the restrictions of both rounds.

\section{NP-Completeness of Two Rounds}\label{p:proof}

We prove NP-hardness for two rounds by a direct reduction from
3-SAT~\cite{Karp72}. Given a formula, we construct two routing trees of
linear size such that
\[
 \boxed{\text{the formula is satisfiable}
 \quad\Longleftrightarrow\quad
 \text{the trees admit two safe update rounds}.}
\]
The trees depend only on the formula; the schedule encodes the assignment.
We use formulas with at least one clause, each containing exactly three
literal positions. Repeated literals are allowed, so clauses of one or
two literals can be padded by repetition without changing satisfiability.

\Needspace{6\baselineskip}
\subsection{Proof Overview}\label{p:overview}

\paragraph{Consistent literal choices.}
A formula is satisfiable exactly when we can choose at least one literal
occurrence in every clause without choosing both $v$ and $\neg v$.
A satisfying assignment provides such choices through its true literals.
Conversely, consistent choices can all be made true and extended to a
full assignment.

\paragraph{Encoding the choices.}
Each occurrence has a \emph{choice router}; assigning it to round~1
selects the occurrence. A \emph{blocker} for each signed literal controls
which occurrences can be selected. Three types of cycles enforce the
following constraints:
\begin{enumerate}
\item A choice and its blocker cannot both be early.
\item At least one blocker of each variable is early.
\item Every clause has an early choice.
\end{enumerate}
The first two constraints enforce consistency. Choosing $v$ makes its
blocker late; the blocker of $\neg v$ is hence early and excludes every
choice of $\neg v$. The third constraint covers every clause.

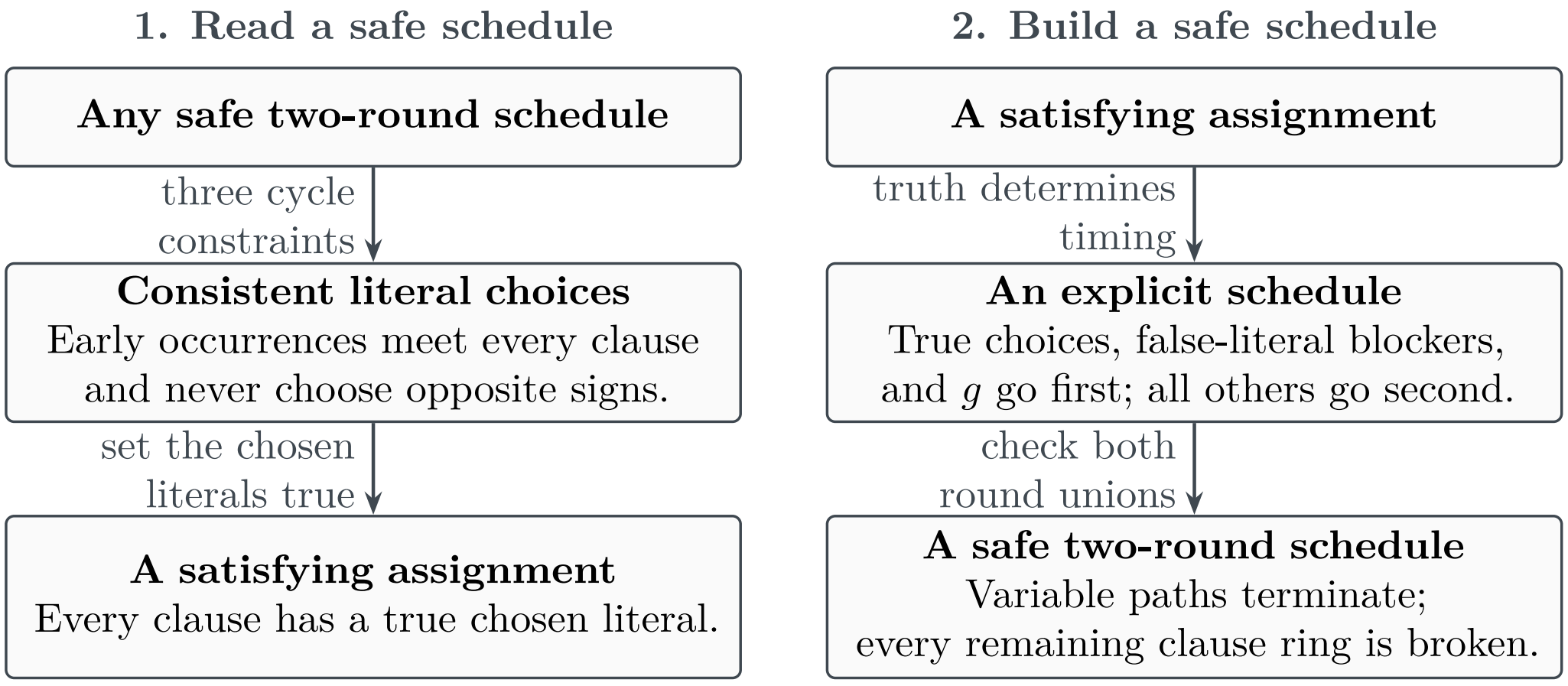
\begin{figure}[!htbp]
\centering
\begin{tikzpicture}[x=1cm,y=1cm,
  proofbox/.style={draw=inkgrey,rounded corners=2pt,line width=.65pt,
    fill=black!2,text width=6.0cm,align=center,inner xsep=5pt,
    inner ysep=3pt,font=\small},
  proofarrow/.style={-{Stealth[length=2mm]},line width=.85pt,inkgrey},
  proofstep/.style={font=\small,text=inkgrey,text width=2.85cm,
    align=right,inner sep=0pt}]
  \node[figheading] at (3.15,.8) {1. Read a safe schedule};
  \node[figheading] at (10.25,.8) {2. Build a safe schedule};

  \node[proofbox,minimum height=.85cm] (schedule) at (3.15,0)
    {\textbf{Any safe two-round schedule}};
  \node[proofbox,minimum height=.85cm] (assignment) at (10.25,0)
    {\textbf{A satisfying assignment}};

  \node[proofbox,minimum height=1.3cm] (choices) at (3.15,-1.95)
    {\textbf{Consistent literal choices}\\
     Early occurrences meet every clause\\
     and never choose opposite signs.};
  \node[proofbox,minimum height=1.3cm] (timing) at (10.25,-1.95)
    {\textbf{An explicit schedule}\\
     True choices, false-literal blockers,\\
     and $g$ go first; all others go second.};

  \node[proofbox,minimum height=1.4cm] (satisfying) at (3.15,-4.15)
    {\textbf{A satisfying assignment}\\
     Every clause has a true chosen literal.};
  \node[proofbox,minimum height=1.4cm] (safe) at (10.25,-4.15)
    {\textbf{A safe two-round schedule}\\
     Variable paths terminate;\\
     every remaining clause ring is broken.};

  \draw[proofarrow] (schedule.south) -- (choices.north)
    node[proofstep,midway,left=4pt] {three cycle\\constraints};
  \draw[proofarrow] (choices.south) -- (satisfying.north)
    node[proofstep,midway,left=4pt] {set the chosen\\literals true};
  \draw[proofarrow] (assignment.south) -- (timing.north)
    node[proofstep,midway,left=4pt] {truth determines\\timing};
  \draw[proofarrow] (timing.south) -- (safe.north)
    node[proofstep,midway,left=4pt] {check both\\round unions};
\end{tikzpicture}
\caption{The two directions of the reduction. The left column extracts consistent
literal choices from an arbitrary safe two-round schedule. The right
column constructs a schedule from a satisfying assignment and proves its safety.
Arrows in this overview are logical steps, not forwarding rules.}
\label{p:roadmap}
\end{figure}

\paragraph{Correctness.}
Every safe two-round schedule obeys these constraints and gives a
satisfying assignment. Conversely, update all true-literal choices,
false-literal blockers, and the shared head~$g$ first; all others go second.
We prove that both union graphs are acyclic: variable branches terminate,
and each clause ring is broken by an early choice.
Figure~\ref{p:roadmap} summarizes the two directions.

\FloatBarrier
\Needspace{6\baselineskip}
\subsection{Construction of the Two Routing Trees}\label{p:construction}

\paragraph{Running example.}
The formula
\[
 \varphi=(x\lor\neg y\lor z)\land(\neg x\lor y\lor z)
\]
is satisfied by $x=\mathrm{true}$, $y=\mathrm{false}$, and
$z=\mathrm{true}$. The first clause has three true occurrences; the second
has only one. The true occurrences become the early choices.

Consider a nonempty formula with $m$ three-position clauses and $n$ appearing
variables. An \emph{occurrence} is one literal position in one clause.
Positions remain distinct even when their literals agree; tautological
clauses are allowed. The construction uses the following routers:
\begin{itemize}
\item A \emph{choice router} $R_{i,j}$ for position $j$ of clause $i$.
      Updating it early selects that occurrence.
\item A \emph{clause closer} $C_i$. Its new edge can close a cycle in
      round~2 if all three choices of its clause are late.
\item For each variable $v$, two \emph{blockers} $B_v,B_{\neg v}$ and two
      \emph{connector routers} $Z_v,Z_{\neg v}$. An early blocker forbids
      early choices of its literal. The connectors couple the two
      opposite blockers.
\item A shared \emph{head} $g$ and the destination $d$. The head starts
      the old clause path and is the new next hop of every blocker.
\end{itemize}
All these routers are distinct. Both signs of every appearing variable
receive their $B,Z$ routers, even if one sign never occurs in a clause.

\begin{figure}[!htbp]
\centering
\begin{tikzpicture}[x=1cm,y=1cm,
  oldpath/.style={oldedge,densely dotted}]
  \fill[black!3,rounded corners=3pt] (1.15,-.57) rectangle (8.05,.57);
  \node[figheading] at (4.6,.95) {One old spine through all clauses};
  \node[auxrouter] (g) at (0,0) {$g$};
  \node[routevertex] (r1) at (1.8,0) {$R_{1,1}$};
  \node[routevertex] (r2) at (3.6,0) {$R_{1,2}$};
  \node[routevertex] (r3) at (5.4,0) {$R_{1,3}$};
  \node[auxrouter] (c) at (7.4,0) {$C_1$};
  \node[destination] (d) at (10.2,-2.0) {$d$};
  \draw[oldedge] (g) -- (r1);
  \draw[oldedge] (r1) -- (r2);
  \draw[oldedge] (r2) -- (r3);
  \draw[oldedge] (r3) -- (c);
  \draw[oldpath] (c) to[out=0,in=90]
    node[figlabel,align=center,above right=2pt] {remaining\\clauses} (d);
  \node[figlabel] at (4.6,-.87) {clause 1};

  \node[figheading] at (2.1,-1.45) {Two old branches for each variable $v$};
  \node[auxrouter] (bp) at (.5,-2.25) {$B_v$};
  \node[auxrouter] (zp) at (3.3,-2.25) {$Z_v$};
  \node[auxrouter] (bn) at (.5,-3.55) {$B_{\neg v}$};
  \node[auxrouter] (zn) at (3.3,-3.55) {$Z_{\neg v}$};
  \draw[oldedge] (bp) -- (zp);
  \draw[oldedge] (zp) -- (d);
  \draw[oldedge] (bn) -- (zn);
  \draw[oldedge] (zn) -- (d);
\end{tikzpicture}
\caption{Old rules for the first clause and one variable.
The dotted blue arrow abbreviates the remaining clause blocks.
Every old path ends at $d$. Repeated literals still have distinct choice
routers. Throughout the proof, solid blue arrows are old rules and
dashed orange arrows are new rules.}
\label{p:old-tree}
\end{figure}
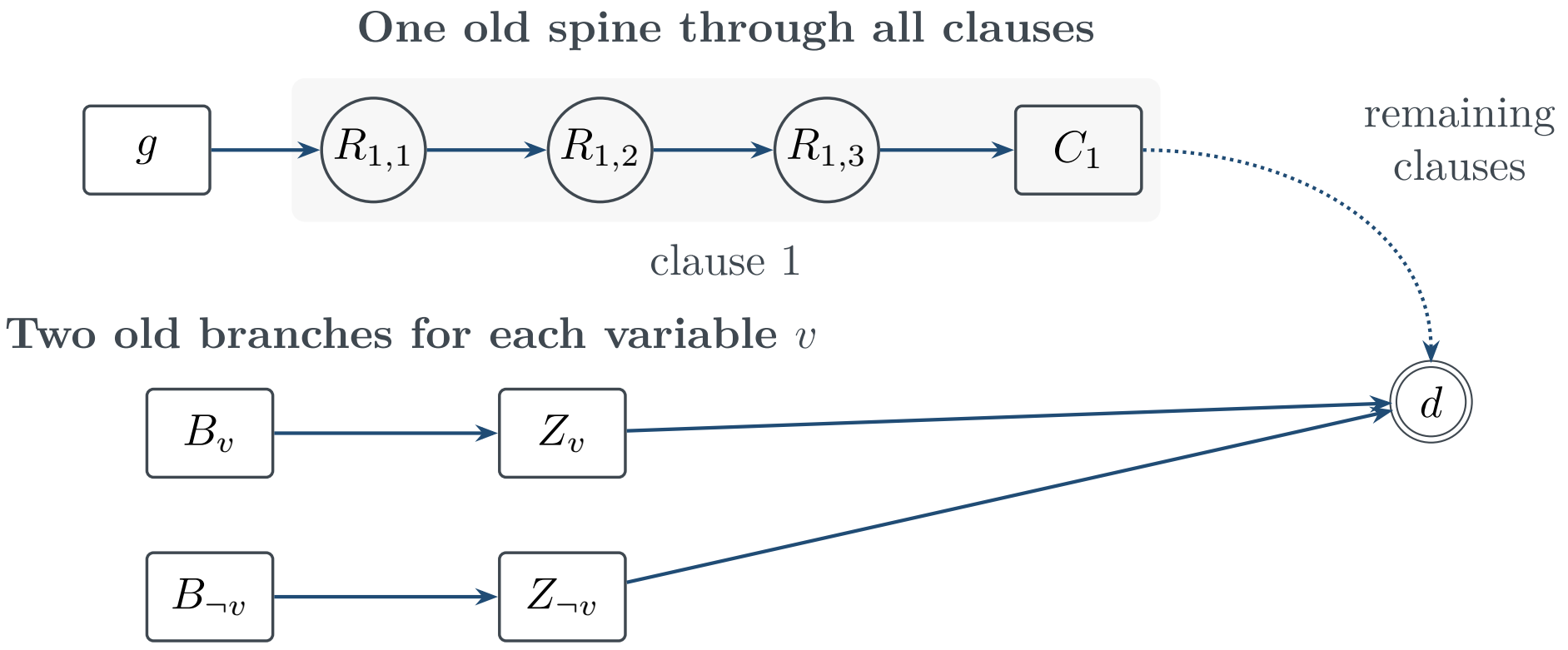

\paragraph{Old rules.}
All choices and closers lie on one path, the \emph{spine}:
\[
 g\to R_{1,1}\to R_{1,2}\to R_{1,3}\to C_1
 \to\cdots\to R_{m,1}\to R_{m,2}\to R_{m,3}\to C_m\to d.
\]
The remaining old paths are $B_\ell\to Z_\ell\to d$, one for each
signed literal $\ell$. Thus the old tree consists of a clause spine and
two short branches per variable (Figure~\ref{p:old-tree}).

\FloatBarrier
\Needspace{4\baselineskip}
\paragraph{New rules.}
A choice points to the blocker of its literal. A closer points back to
its own clause's first choice. Every blocker points to $g$, and $g$
points to $d$. Finally, each connector points to the opposite blocker.
Table~\ref{p:rules} specifies the complete construction. Here $\ell$ is
the literal at the indicated occurrence, and $\neg\ell$ denotes its
opposite sign. The destination has no outgoing rule.

\begin{table}[!htbp]
\centering
\caption{The complete successor table. There is one choice per literal
position, one closer per clause, and four routers per variable.}
\label{p:rules}
\begin{tabularx}{\linewidth}{@{}lXX@{}}
\toprule
Router & Old successor & New successor \\
\midrule
$R_{i,j}$ & $R_{i,j+1}$ if $j<3$; $C_i$ if $j=3$ & $B_\ell$ \\
$C_i$ & $R_{i+1,1}$ if $i<m$; $d$ if $i=m$ & $R_{i,1}$ \\
$B_\ell$ & $Z_\ell$ & $g$ \\
$Z_\ell$ & $d$ & $B_{\neg\ell}$ \\
$g$ & $R_{1,1}$ & $d$ \\
\bottomrule
\end{tabularx}
\end{table}

\begin{figure}[!htbp]
\centering
\begin{tikzpicture}[x=1cm,y=1cm]
  \node[figheading] at (6.0,1.8)
    {The new rules collect all paths at $g$, then $d$};

  \node[auxrouter] (ci) at (0,.4) {$C_i$};
  \node[routevertex] (ri) at (2.6,.4) {$R_{i,1}$};
  \node[routevertex] (rj) at (2.6,-1.0) {$R_{j,k}$};
  \node[routevertex] (rn) at (2.6,-2.5) {$R_{p,q}$};
  \node[figlabel,anchor=north] at ([yshift=-2mm]ri.south) {literal $v$};
  \node[figlabel,anchor=north] at ([yshift=-2mm]rj.south) {literal $v$};
  \node[figlabel,anchor=north] at ([yshift=-2mm]rn.south) {literal $\neg v$};

  \node[auxrouter] (bp) at (6.1,0) {$B_v$};
  \node[auxrouter] (bn) at (6.1,-2.5) {$B_{\neg v}$};
  \node[auxrouter] (zn) at (4.4,1.0) {$Z_{\neg v}$};
  \node[auxrouter] (zp) at (4.4,-3.35) {$Z_v$};
  \node[auxrouter] (g) at (9.2,-1.2) {$g$};
  \node[destination] (d) at (12.0,-1.2) {$d$};

  \draw[newedge] (ci) -- (ri);
  \draw[newedge] (ri) -- (bp);
  \draw[newedge] (rj) -- (bp);
  \draw[newedge] (rn) -- (bn);
  \draw[newedge] (zn) -- (bp);
  \draw[newedge] (zp) -- (bn);
  \draw[newedge] (bp) -- (g);
  \draw[newedge] (bn) -- (g);
  \draw[newedge] (g) -- (d);
\end{tikzpicture}
\caption{The new rules, shown on a representative part of the construction.
Choices carrying the same literal share a blocker; the two connectors
point to the opposite blockers. These branches merge at a blocker and continue to
$g$ and $d$. A closer adds one preceding edge, so the maximum new-path
length is four. Only the indicated routers are shown.}
\label{p:new-tree}
\end{figure}
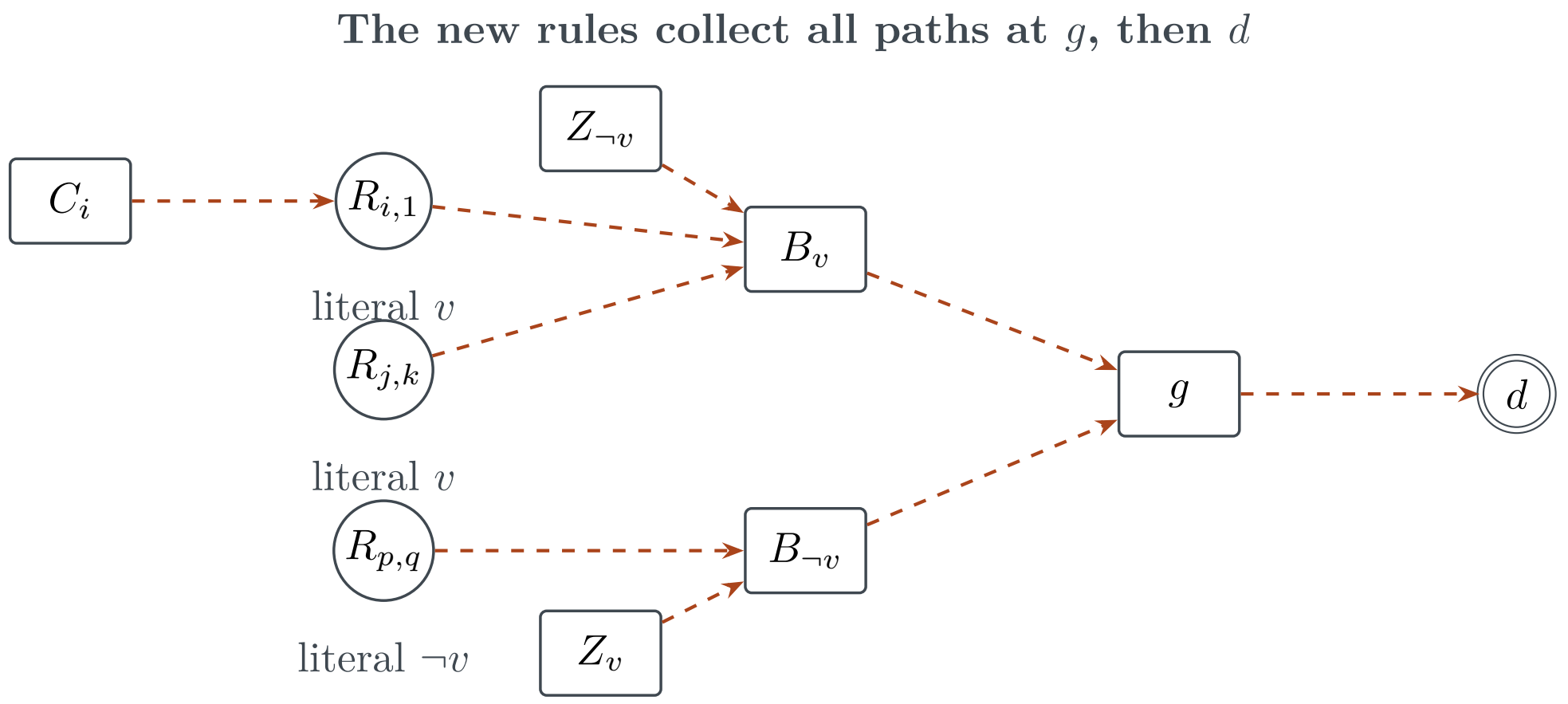

Figure~\ref{p:new-tree} shows how the new paths merge toward the destination.

\FloatBarrier
\paragraph{Both endpoints are trees.}
All old paths end at $d$. New paths are suffixes of
$C\to R\to B\to g\to d$ or $Z\to B\to g\to d$; they reach $d$
in at most four edges. Each nonroot has exactly one successor in either
endpoint, so both graphs are spanning routing trees. Every nonroot rule
changes: in particular, $g$ changes from $R_{1,1}$ to $d$ because
$m\geq1$. The construction has $3m+m+4n+2=4n+4m+2$ routers and is linear
in the formula size.

\Needspace{6\baselineskip}
\subsection{From Two Safe Rounds to a Satisfying Assignment}\label{p:necessary}

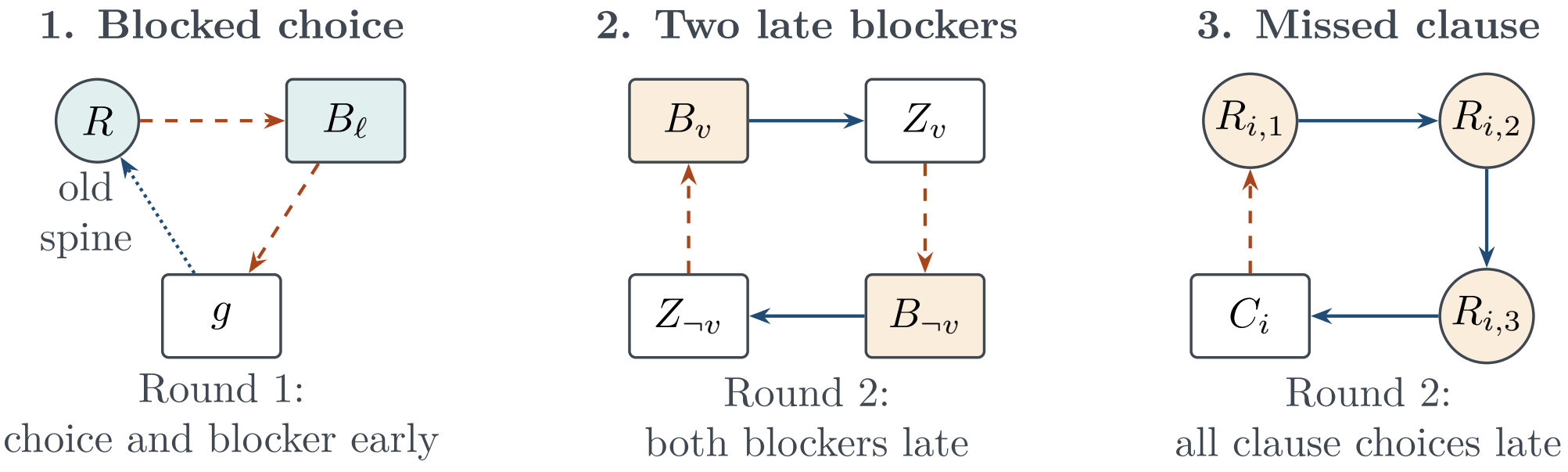
\begin{figure}[!htbp]
\centering
\begin{tikzpicture}[x=1cm,y=1cm,
  oldpath/.style={oldedge,densely dotted}]
  \node[figheading] at (1.05,.82) {1. Blocked choice};
  \node[routevertex,fill=earlyfill] (r) at (0,0) {$R$};
  \node[auxrouter,fill=earlyfill] (b) at (2.1,0) {$B_\ell$};
  \node[auxrouter] (g) at (1.05,-1.65) {$g$};
  \draw[newedge] (r) -- (b);
  \draw[newedge] (b) -- (g);
  \draw[oldpath] (g) -- node[figlabel,left=2pt,align=center] {old\\spine} (r);
  \node[figlabel,align=center] at (1.05,-2.5) {Round 1:\\choice and blocker early};

  \node[figheading] at (6.0,.82) {2. Two late blockers};
  \node[auxrouter,fill=latefill] (bp) at (5,0) {$B_v$};
  \node[auxrouter] (zp) at (7,0) {$Z_v$};
  \node[auxrouter,fill=latefill] (bn) at (7,-1.65) {$B_{\neg v}$};
  \node[auxrouter] (zn) at (5,-1.65) {$Z_{\neg v}$};
  \draw[oldedge] (bp) -- (zp);
  \draw[newedge] (zp) -- (bn);
  \draw[oldedge] (bn) -- (zn);
  \draw[newedge] (zn) -- (bp);
  \node[figlabel,align=center] at (6,-2.5) {Round 2:\\both blockers late};

  \node[figheading] at (10.75,.82) {3. Missed clause};
  \node[routevertex,fill=latefill] (r1) at (9.75,0) {$R_{i,1}$};
  \node[routevertex,fill=latefill] (r2) at (11.75,0) {$R_{i,2}$};
  \node[routevertex,fill=latefill] (r3) at (11.75,-1.65) {$R_{i,3}$};
  \node[auxrouter] (c) at (9.75,-1.65) {$C_i$};
  \draw[oldedge] (r1) -- (r2);
  \draw[oldedge] (r2) -- (r3);
  \draw[oldedge] (r3) -- (c);
  \draw[newedge] (c) -- (r1);
  \node[figlabel,align=center] at (10.75,-2.5) {Round 2:\\all clause choices late};
\end{tikzpicture}
\caption{The three cycle constraints. Teal routers are early; tan
routers are late. Uncolored routers may be in either round. Panel~1 is a
cycle in $G_1$; panels~2 and~3 are cycles in $G_2$. The dotted edge is an
old spine path. By Lemma~\ref{p:union}, each cycle occurs in some actual
partial completion.}
\label{p:obstructions}
\end{figure}

Consider an arbitrary safe two-round schedule and select exactly the occurrences
whose $R$-routers are early. The three cycles in
Figure~\ref{p:obstructions} impose the following constraints.
Since \emph{all old edges are available in $G_1$} and
\emph{all new edges are available in $G_2$}, the argument holds regardless
of the rounds assigned to $g$, any $C_i$, or any $Z_\ell$.

\begin{enumerate}
\item \textbf{An early choice cannot have an early blocker.}
Suppose $R$ carries literal $\ell$ and both $R$ and $B_\ell$ are early.
Their new edges $R\to B_\ell\to g$ and the old spine from $g$ back
to $R$ form a cycle in $G_1$. Hence $B_\ell$ must be late whenever an
occurrence of $\ell$ is chosen.
\item \textbf{At least one blocker of each variable is early.}
If both $B_v$ and $B_{\neg v}$ are late, their old edges remain in
$G_2$. Together with the connectors' new edges, they form the cycle
\[
 B_v\xrightarrow{\rm old}Z_v\xrightarrow{\rm new}B_{\neg v}
 \xrightarrow{\rm old}Z_{\neg v}\xrightarrow{\rm new}B_v.
\]
\item \textbf{Every clause has an early choice.}
If $R_{i,1},R_{i,2},R_{i,3}$ are all late, $G_2$ contains their old
edges along the clause and the closer's new return edge:
\[
 R_{i,1}\xrightarrow{\rm old}R_{i,2}\xrightarrow{\rm old}R_{i,3}
 \xrightarrow{\rm old}C_i\xrightarrow{\rm new}R_{i,1}.
\]
This contradicts safety of the second round.
\end{enumerate}

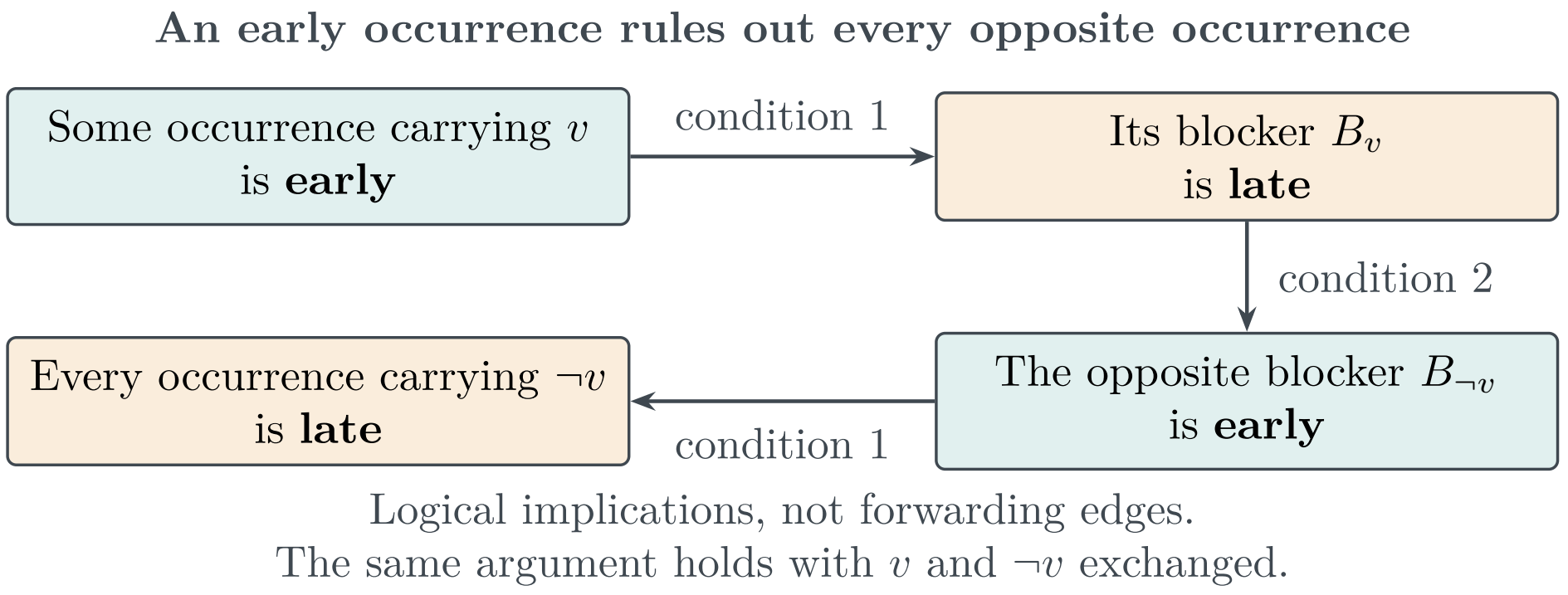
\begin{figure}[!htbp]
\centering
\begin{tikzpicture}[x=1cm,y=1cm,
  logicbox/.style={draw=inkgrey,rounded corners=2pt,line width=.65pt,
    text width=4.6cm,minimum height=1cm,align=center,inner sep=5pt,font=\small},
  implies/.style={-{Stealth[length=2mm]},line width=.85pt,inkgrey}]
  \node[figheading] at (6.3,1.0)
    {An early occurrence rules out every opposite occurrence};

  \node[logicbox,fill=earlyfill] (positive) at (2.6,0)
    {Some occurrence carrying $v$\\is \textbf{early}};
  \node[logicbox,fill=latefill] (positiveblocker) at (10.0,0)
    {Its blocker $B_v$\\is \textbf{late}};
  \node[logicbox,fill=earlyfill] (negativeblocker) at (10.0,-1.95)
    {The opposite blocker $B_{\neg v}$\\is \textbf{early}};
  \node[logicbox,fill=latefill] (negative) at (2.6,-1.95)
    {Every occurrence carrying $\neg v$\\is \textbf{late}};

  \draw[implies] (positive.east) -- (positiveblocker.west)
    node[figlabel,midway,above=2pt] {condition 1};
  \draw[implies] (positiveblocker.south) -- (negativeblocker.north)
    node[figlabel,midway,right=3pt] {condition 2};
  \draw[implies] (negativeblocker.west) -- (negative.east)
    node[figlabel,midway,below=2pt] {condition 1};

  \node[figlabel,align=center] at (6.3,-3.05)
    {Logical implications, not forwarding edges.\\
     The same argument holds with $v$ and $\neg v$ exchanged.};
\end{tikzpicture}
\caption{Literal consistency from the first two cycle constraints.
The same implications hold with $v$ and $\neg v$ exchanged. The arrows
are logical implications valid for every safe two-round schedule;
they do not require specified rounds for $g$, the closers, or the
connectors. Other occurrences of $v$ need not be early.}
\label{p:literal-consistency}
\end{figure}

\FloatBarrier
\Needspace{4\baselineskip}
\paragraph{Extracting the assignment.}
If an occurrence of $v$ is chosen, its blocker must be late. The opposite
blocker is then early and forces every occurrence of $\neg v$ late
(Figure~\ref{p:literal-consistency}). Opposite literals hence cannot both
be chosen.
Set $v$ true if a chosen occurrence carries $v$,
and false if one carries $\neg v$; if neither sign was chosen, assign
$v$ arbitrarily. Requirement~3 supplies a true literal in every clause.
Thus the formula is satisfiable.

\FloatBarrier
A clause may have several early choices. A variable may have both blockers
early if neither sign is chosen. Only chosen occurrences determine the
assignment; a late occurrence may still represent a true literal.

\Needspace{6\baselineskip}
\subsection{From a Satisfying Assignment to Two Safe Rounds}\label{p:sufficient}

Conversely, fix a satisfying assignment. We update $g$ in round~1 and
every closer $C_i$ and connector $Z_\ell$ in round~2. The choices and
blockers follow Table~\ref{p:assignment-schedule}.

\begin{table}[!htbp]
\centering
\caption{The schedule induced by a satisfying assignment. A true literal has
early choices but a late blocker; a false literal has the opposite timing.}
\label{p:assignment-schedule}
\begin{tabularx}{\linewidth}{@{}lXX@{}}
\toprule
Value of literal $\ell$ & Every choice carrying $\ell$ & Blocker $B_\ell$ \\
\midrule
True & Round 1 (early) & Round 2 (late) \\
False & Round 2 (late) & Round 1 (early) \\
\bottomrule
\end{tabularx}
\end{table}

It remains to prove that both round unions are acyclic. The following
path arguments explain why; Appendix~\ref{app:ranks} gives explicit
integer-valued certificates that strictly decrease on every edge.

\Needspace{4\baselineskip}
\paragraph{Round 1: paths leaving the spine terminate.}
All connectors and closers are late, so they have only old edges in
$G_1$. A walk starting on the spine follows old edges toward $d$ or leaves
through the new edge of an early choice $R$. Such an $R$ carries a true literal $\ell$,
so its blocker is late. The new edge therefore leads along
\[
 R\xrightarrow{\rm new}B_\ell
   \xrightarrow{\rm old}Z_\ell\xrightarrow{\rm old}d.
\]
It cannot return to the spine. A connector leads directly to $d$.
A blocker either follows its old branch to $d$ or enters the spine at
$g$, where the walk takes the new edge to $d$ or follows the terminating
spine paths just described. Thus every walk terminates, and $G_1$ is acyclic.

\begin{figure}[!htbp]
\centering
\begin{tikzpicture}[x=1cm,y=1cm]
  \node[figheading,anchor=west] at (0,-.05)
    {Round 1: an early true occurrence enters a fixed old path};
  \node[routevertex,fill=earlyfill] (r1) at (.6,-.85) {$R$};
  \node[auxrouter,fill=latefill] (b1) at (3.4,-.85) {$B_v$};
  \node[auxrouter,fill=latefill] (z1) at (6.2,-.85) {$Z_v$};
  \node[destination] (d1) at (9,-.85) {$d$};
  \draw[newedge] (r1) -- (b1);
  \draw[oldedge] (b1) -- (z1);
  \draw[oldedge] (z1) -- (d1);
  \node[figlabel] at (.6,-1.48) {early ($v$)};
  \node[figlabel] at (3.4,-1.48) {late};
  \node[figlabel] at (6.2,-1.48) {late};

  \node[figheading,anchor=west] at (0,-2.4)
    {Round 2: even the remaining old blocker edge leads to $d$};
  \node[auxrouter,fill=latefill] (b2) at (.6,-3.65) {$B_v$};
  \node[auxrouter,fill=latefill] (z2) at (3.4,-3.65) {$Z_v$};
  \node[auxrouter,fill=earlyfill] (bn2) at (6.2,-3.65) {$B_{\neg v}$};
  \node[auxrouter,fill=earlyfill] (g2) at (9,-3.65) {$g$};
  \node[destination] (d2) at (11.8,-3.65) {$d$};
  \draw[oldedge] (b2) -- (z2);
  \draw[newedge] (z2) -- (bn2);
  \draw[newedge] (bn2) -- (g2);
  \draw[newedge] (g2) -- (d2);
  \draw[newedge] (b2.north) -- (.6,-2.9) -- (9,-2.9) -- (g2.north);
  \node[figlabel,fill=white,inner sep=2pt] at (4.8,-2.9)
    {or use $B_v$'s new edge};
  \draw[oldedge] (z2.south east) -- (4.4,-4.65) -- (11.8,-4.65) -- (d2.south);
  \node[figlabel,fill=white,inner sep=2pt] at (7.6,-4.65)
    {or use $Z_v$'s old edge};
  \node[figlabel] at (.6,-4.25) {late};
  \node[figlabel] at (3.4,-4.25) {late};
  \node[figlabel] at (6.2,-4.25) {early};
  \node[figlabel] at (9,-4.25) {early};
\end{tikzpicture}
\caption{Terminating variable paths under the assignment-derived
schedule, shown for $v=\mathrm{true}$. The top panel follows an early
choice after it switches. The bottom panel includes both options at the
late blocker and its connector. These are selected terminating paths,
not drawings of the complete union graphs.}
\label{p:variable-schedule}
\end{figure}
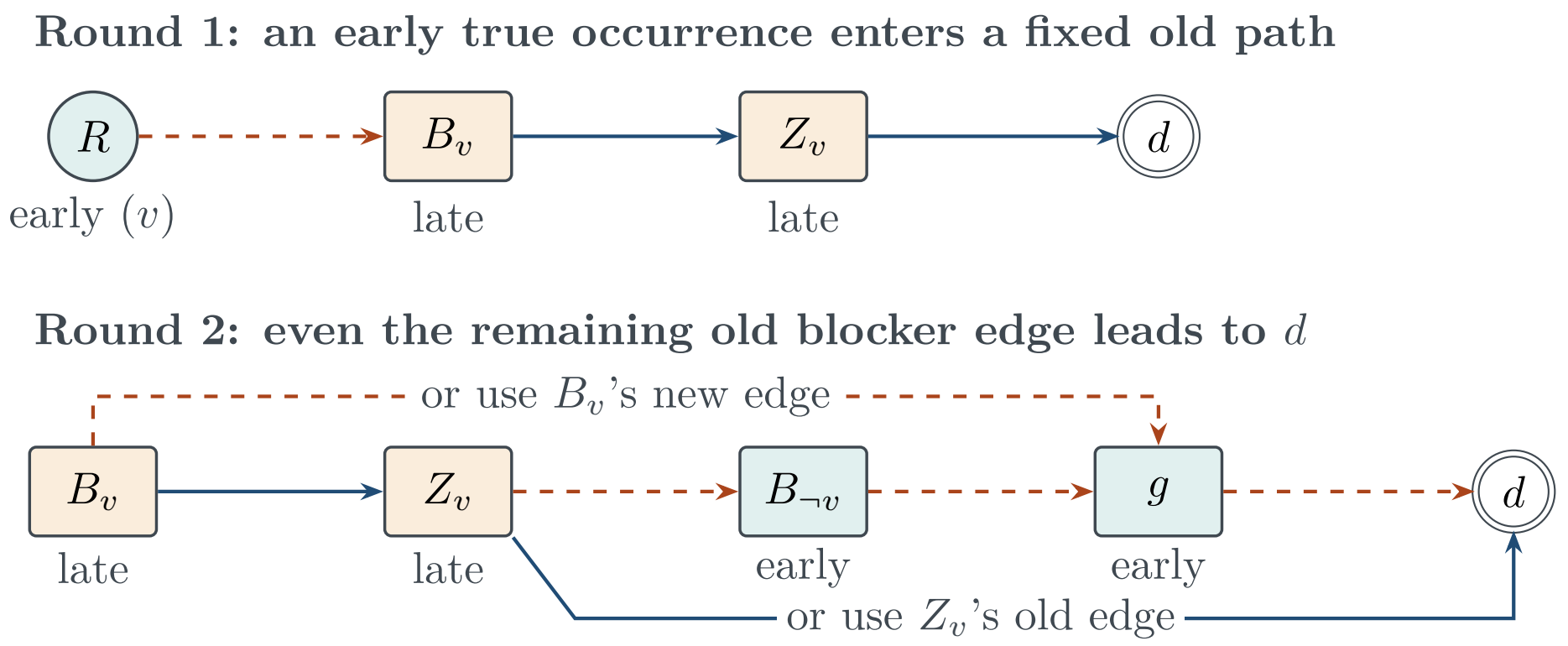

\paragraph{Round 2: variable branches terminate.}
The early head $g$ now has only its new edge to $d$. A false-literal
blocker is early and therefore goes straight to $g$. A true-literal
blocker is late: it can use its new edge to $g$, or its old edge to
$Z_\ell$. That connector can go to $d$ by its old rule or to the
opposite, early blocker by its new rule. Every option terminates.
The connector of a false literal points either to $d$ or to the opposite,
true-literal blocker. Both continuations terminate.
Hence a cycle cannot use a variable branch or a choice's new edge
(Figure~\ref{p:variable-schedule}).

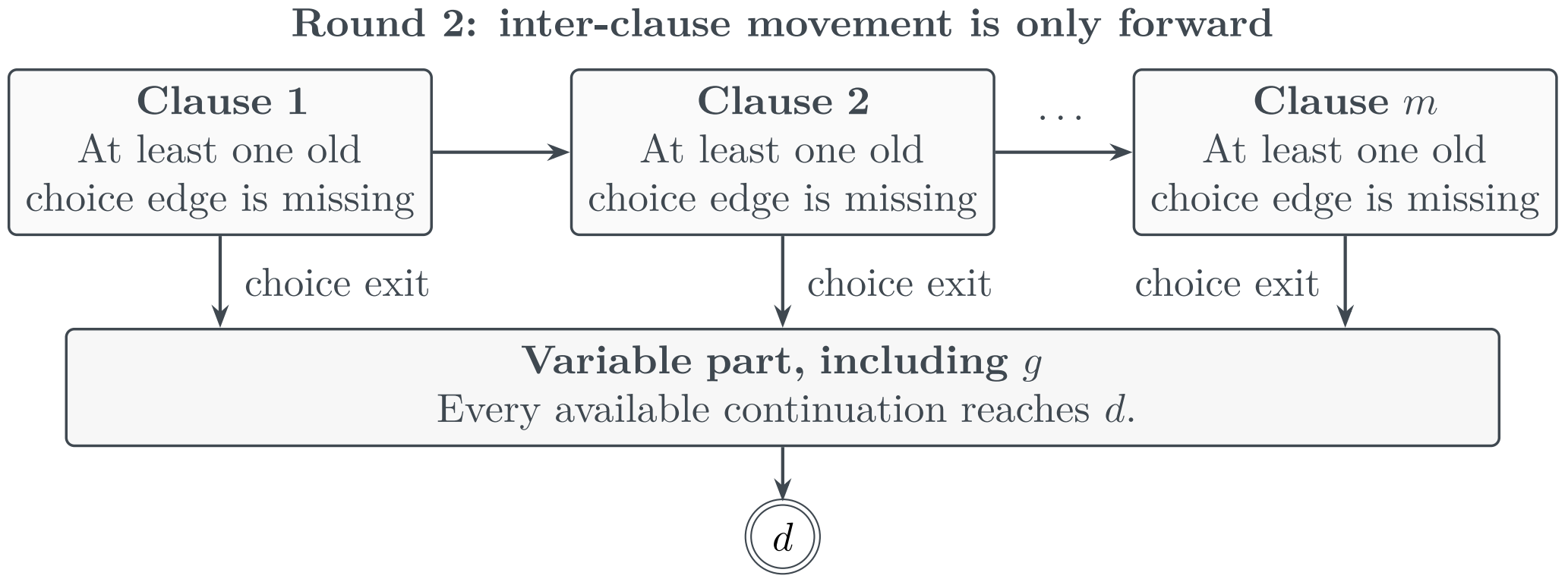
\begin{figure}[!htbp]
\centering
\begin{tikzpicture}[x=1cm,y=1cm,
  clauseblock/.style={draw=inkgrey,rounded corners=2pt,line width=.65pt,
    fill=black!2,text width=3.4cm,minimum height=1.25cm,
    inner sep=4pt,align=center,font=\small,text=inkgrey},
  groupflow/.style={-{Stealth[length=2mm]},line width=.8pt,inkgrey}]
  \node[figheading,align=center] at (7.1,.45)
    {Round 2: inter-clause movement is only forward};

  \node[clauseblock] (c1) at (2.2,-.65)
    {\textbf{Clause 1}\\
     At least one old\\choice edge is missing};
  \node[clauseblock] (c2) at (7.1,-.65)
    {\textbf{Clause 2}\\
     At least one old\\choice edge is missing};
  \node[clauseblock] (cm) at (12,-.65)
    {\textbf{Clause $m$}\\
     At least one old\\choice edge is missing};
  \draw[groupflow] (c1.east) -- (c2.west);
  \draw[groupflow] (c2.east) --
    node[figlabel,above=2pt] {$\cdots$} (cm.west);

  \node[draw=inkgrey,rounded corners=2pt,line width=.65pt,
    fill=black!3,text width=12.2cm,minimum height=.85cm,
    inner sep=4pt,align=center,font=\small,text=inkgrey] (part) at (7.1,-2.7)
    {\textbf{Variable part, including $g$}\\
     Every available continuation reaches $d$.};
  \draw[groupflow] (c1.south) --
    node[figlabel,right=2pt] {choice exit} (part.north -| c1.south);
  \draw[groupflow] (c2.south) --
    node[figlabel,right=2pt] {choice exit} (part.north -| c2.south);
  \draw[groupflow] (cm.south) --
    node[figlabel,left=2pt] {choice exit} (part.north -| cm.south);
  \node[destination] (dest) at (7.1,-4.0) {$d$};
  \draw[groupflow] (part.south) -- (dest.north);

\end{tikzpicture}
\caption{Second-round paths under the satisfying-assignment schedule.
Boxes group routers. Movement between clauses is only forward, and every
exit terminates at $d$. A cycle could therefore only be a local clause
ring, broken by an early choice (Figure~\ref{p:choice-example}). Direct
exits to $d$ are omitted; with fewer clauses, use fewer boxes.}
\label{p:clause-order}
\end{figure}

\FloatBarrier
\Needspace{4\baselineskip}
\paragraph{Round 2: every remaining clause ring is broken.}
Any remaining cycle must stay on the spine. Within a
clause, old choice edges move forward. A closer can use its old edge
to the next clause (or $d$), or its new edge back to its own first
choice. No edge goes back to an earlier clause
(Figure~\ref{p:clause-order}). A cycle must therefore stay inside one
clause and traverse all three old choice edges. At least one literal of
that clause is true under the assignment.
Its choice is early, so its old edge is absent from $G_2$.
The clause ring is hence broken. Thus $G_2$ is acyclic, and
Lemma~\ref{p:union} gives safety throughout both rounds.

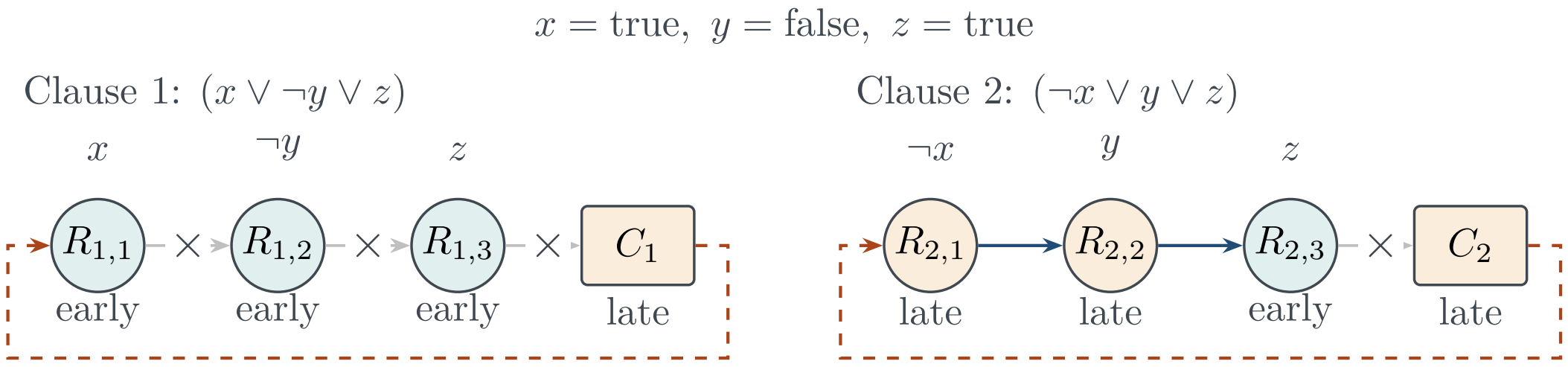
\begin{figure}[!htbp]
\centering
\begin{tikzpicture}[x=1cm,y=1cm,
  removed/.style={-{Stealth[length=1.8mm]},line width=.7pt,black!25},
  cutmark/.style={font=\large\bfseries,text=inkgrey,fill=white,inner sep=1pt}]
  \node[figlabel,anchor=center] at (6.9,1.0)
    {$x=\mathrm{true},\ y=\mathrm{false},\ z=\mathrm{true}$};

  \begin{scope}
  \node[figlabel,anchor=west] at (0,.4) {Clause 1: $(x\lor\neg y\lor z)$};
  \node[routevertex,fill=earlyfill] (r11) at (.8,-.95) {$R_{1,1}$};
  \node[routevertex,fill=earlyfill] (r12) at (2.4,-.95) {$R_{1,2}$};
  \node[routevertex,fill=earlyfill] (r13) at (4.0,-.95) {$R_{1,3}$};
  \node[auxrouter,fill=latefill] (c1) at (5.6,-.95) {$C_1$};
  \node[figlabel,anchor=south] at ([yshift=2mm]r11.north) {$x$};
  \node[figlabel,anchor=south] at ([yshift=2mm]r12.north) {$\neg y$};
  \node[figlabel,anchor=south] at ([yshift=2mm]r13.north) {$z$};
  \draw[removed] (r11) -- (r12);
  \draw[removed] (r12) -- (r13);
  \draw[removed] (r13) -- (c1);
  \node[cutmark] at (1.6,-.95) {$\times$};
  \node[cutmark] at (3.2,-.95) {$\times$};
  \node[cutmark] at (4.8,-.95) {$\times$};
  \draw[newedge] (c1.east) -- (6.4,-.95) -- (6.4,-1.95)
    -- (0,-1.95) -- (0,-.95) -- (r11.west);
  \node[figlabel] at (.8,-1.53) {early};
  \node[figlabel] at (2.4,-1.53) {early};
  \node[figlabel] at (4.0,-1.53) {early};
  \node[figlabel] at (5.6,-1.53) {late};
  \end{scope}

  \begin{scope}[shift={(7.4,0)}]
  \node[figlabel,anchor=west] at (0,.4) {Clause 2: $(\neg x\lor y\lor z)$};
  \node[routevertex,fill=latefill] (r21) at (.8,-.95) {$R_{2,1}$};
  \node[routevertex,fill=latefill] (r22) at (2.4,-.95) {$R_{2,2}$};
  \node[routevertex,fill=earlyfill] (r23) at (4.0,-.95) {$R_{2,3}$};
  \node[auxrouter,fill=latefill] (c2) at (5.6,-.95) {$C_2$};
  \node[figlabel,anchor=south] at ([yshift=2mm]r21.north) {$\neg x$};
  \node[figlabel,anchor=south] at ([yshift=2mm]r22.north) {$y$};
  \node[figlabel,anchor=south] at ([yshift=2mm]r23.north) {$z$};
  \draw[oldedge] (r21) -- (r22);
  \draw[oldedge] (r22) -- (r23);
  \draw[removed] (r23) -- (c2);
  \node[cutmark] at (4.8,-.95) {$\times$};
  \draw[newedge] (c2.east) -- (6.4,-.95) -- (6.4,-1.95)
    -- (0,-1.95) -- (0,-.95) -- (r21.west);
  \node[figlabel] at (.8,-1.53) {late};
  \node[figlabel] at (2.4,-1.53) {late};
  \node[figlabel] at (4.0,-1.53) {early};
  \node[figlabel] at (5.6,-1.53) {late};
  \end{scope}
\end{tikzpicture}
\caption{The two-clause example in round~2. An early choice has already
removed its old spine edge, marked by a cross. All choices have a new
edge to their blocker; those edges and the forward edges between clauses
are omitted here. One removed edge per ring suffices, even though the
first clause has three.}
\label{p:choice-example}
\end{figure}

\FloatBarrier
\Needspace{4\baselineskip}
\paragraph{The running example.}
For $x=\mathrm{true}$, $y=\mathrm{false}$, and $z=\mathrm{true}$,
the first round is
\[
 S=\{g,\ R_{1,1},R_{1,2},R_{1,3},R_{2,3},\ B_{\neg x},B_y,B_{\neg z}\}.
\]
Every other router is late. The first clause has three early choices;
the second has one. Both occurrences of $z$ are separate choice routers
but use the same blocker $B_z$. Figure~\ref{p:choice-example} shows the
resulting breaks in the clause rings. The absent literal $\neg z$ still
has its blocker and connector, as required by the construction.

\FloatBarrier
\Needspace{6\baselineskip}
\subsection{Three Rounds Suffice for the Construction}\label{p:three}

The following fixed schedule uses three rounds for every formula:
\[
 \underbrace{\{g\}}_{\text{round 1}}\ ;\qquad
 \underbrace{\text{all choices }R\text{ and blockers }B}_{\text{round 2}}\ ;\qquad
 \underbrace{\text{all connectors }Z\text{ and closers }C}_{\text{round 3}}.
\]
Every partial completion is safe:
\begin{enumerate}
\item Updating $g$ only replaces the start of the old spine by a direct
      edge to $d$. Either state is safe.
\item Closers and connectors still use their old rules. A blocker using its old
      rule follows $B\to Z\to d$; one using its new rule follows
      $B\to g\to d$. A choice either continues along the old spine
      or enters one of these terminating branches. Thus a choice and
      its blocker may finish in \emph{either} order.
\item Choices and blockers now use only new rules. Either rule of a
      connector or closer leads to $d$ or to an already terminating
      blocker or choice. This last round is safe as well.
\end{enumerate}

Finally, a clause ring excludes a one-round schedule: its old choice
edges and new closing edge are all present in the one-round union.
The optimum is hence exactly two for satisfiable formulas and exactly
three otherwise.

For the total reduction, prepend a tautological clause to every well-formed
input, including the empty formula, and replace malformed inputs by a fixed
contradiction. The nonempty result preserves membership in 3-SAT, so the
exact gap applies, with $n,m$ measured after preprocessing.

Both satisfying and unsatisfying inputs yield valid tree pairs and the
displayed safe three-round schedule in polynomial time.
This proves NP-hardness for both parts of
Theorem~\ref{thm:main}: omitting the supplied schedule gives part~(a),
and retaining it gives part~(b).

For membership in NP, first check that the successor tables are well
formed and that both endpoints are spanning routing trees toward the
specified destination. Guess a first-round set $S$ and test the two
unions $G_1(S)$ and $G_2(S)$ for acyclicity. For part~(b), also check that
the supplied schedule assigns every non-destination router to exactly
one of its three rounds and test its three round unions for acyclicity.
Lemma~\ref{p:union} justifies all these safety tests. Every check is
polynomial, proving Theorem~\ref{thm:main}.

\section{Consequences and Related Models}
\label{r:models}

\paragraph{An approximation consequence.}
The reduction separates instances with optimum two from instances with
optimum three, while providing a valid three-round schedule in both cases.
The gap gives the following consequence.
\begin{corollary}\label{cor:approx}
Unless $\mathrm{P}=\mathrm{NP}$, the minimum number of strongly
loop-free asynchronous rounds cannot be approximated in polynomial time
within any factor strictly below $3/2$, even when a safe three-round
schedule is supplied.
\end{corollary}
\begin{proof}
An approximation ratio smaller than $3/2$ would return at most two
rounds when the optimum is two, since the output is an integer. When the
optimum is three, every feasible output uses at least three rounds.
Thus it would decide the source 3-SAT formula in polynomial time.
\end{proof}

Earlier three-round hardness~\cite[Theorem~1]{FoersterW16} rules out factors
strictly below $4/3$ by integrality. When a safe three-round schedule is
supplied, factor $3/2$ is attainable:
return zero rounds if $U=\varnothing$, one round if $E_0\cup E_1$ is
acyclic, and the supplied three rounds otherwise. In the last case the
optimum is at least two.

\paragraph{Related update models.}
Related update models vary the endpoint graphs, safety condition,
allowed operations, or objective. Table~\ref{r:comparison} compares them;
see also the survey
of Foerster et al.~\cite[Sections~III and~IV]{FoersterSV19}.

\begin{table}[htbp]
\centering
\small
\caption{Two-round results depend on the endpoint graphs, allowed updates,
safety condition, and optimization objective.}
\label{r:comparison}
\begin{tabularx}{\linewidth}{@{}>{\raggedright\arraybackslash}p{.23\linewidth}
  >{\raggedright\arraybackslash}p{.27\linewidth}
  >{\raggedright\arraybackslash}X@{}}
\toprule
Setting & What changes & Consequence for the comparison\\
\midrule
Two source--destination paths~\cite{FoersterLMS18,LudwigMS15} & Restricted endpoint graphs &
Two rounds are polynomial; three are NP-complete under strong loop freedom.\\
\addlinespace
Relaxed loop freedom~\cite{FoersterLMS18} & Only ingress-reachable cycles are forbidden &
A different safety predicate; for path pairs, two-round feasibility
nevertheless agrees with the strong version.\\
\addlinespace
Maximum safe batch~\cite{AmiriLMS16,FoersterW16} & Only the next round is optimized &
Optimizes $|S|$ subject only to acyclicity of $G_1(S)$; two-round feasibility also requires $G_2(S)$.\\
\addlinespace
Local verification~\cite{FoersterLSW18,FoersterS19} & Routers check labels and apply updates locally &
Certifying safe updates does not minimize the number of global batches.\\
\addlinespace
Fixed-spanning, all-target parent migration~\cite{GaoMGWJKC24} & Arcs are reversed &
Arc reversal yields the analogous abstract all-target reachability problem for atomic parent replacement; additional protocol requirements may differ.\\
\addlinespace
Several destinations or overlapping policies~\cite{VanbeverVPPB12,YuBZFL18,YuFZB17} & Update actions can couple
logical routing rules &
Hardness for coupled rules need not hold for one successor pair per router.\\
\addlinespace
Versioned two-phase updates~\cite{ReitblattFFMW12} & Extra rules and packet tags are available &
Two protocol phases are not two rounds of prescribed replacements.\\
\bottomrule
\end{tabularx}
\end{table}

\paragraph{Paths versus trees.}
For one old and one new source--destination path, two-round feasibility is
polynomial, whereas three-round feasibility is NP-complete under strong
loop freedom~\cite{FoersterLMS18,LudwigMS15}. Tree updates allow several
routers to share a next hop, with different merging patterns in the two
endpoint trees. The path algorithm does not cover these instances.
Three-round hardness for one destination was known
\cite[Section~VI]{FoersterLMS18}\cite[Theorem~1]{FoersterW16}, while
Foerster et al.\ left two rounds open~\cite[Section~VI]{FoersterLMS18}.
Theorem~\ref{thm:main} establishes NP-completeness for two rounds. The polynomial
two-round results for single-destination updates in earlier accounts
concern path pairs~\cite{FoersterSV19,FoersterThesis16}.
For source--destination paths, R\"acke, Schmid, and Vintan give exact
integer-programming algorithms for strong and relaxed loop freedom and a
local search heuristic for the relaxed variant~\cite[Sections~III and~IV-C]{RackeSV24}.

\paragraph{Strong and relaxed loop freedom.}
Strong and relaxed loop freedom can both require safety under every partial
completion. Strong loop freedom forbids cycles throughout the graph;
the relaxed variant forbids only cycles reachable from a designated
ingress and can therefore admit a cycle isolated from that ingress. For path pairs,
the two variants nevertheless have identical
two-round yes-instances~\cite[Section~IV-A, Observation~1]{FoersterLMS18}.
That statement is restricted to paths; our tree theorem does not classify
two-round feasibility under a chosen-ingress-only requirement.

\paragraph{Maximum batch size versus completion time.}
Computing a maximum-size safe batch is NP-hard for path pairs, as shown
by Amiri et al.~\cite{AmiriLMS16}, and for one-destination trees, as shown
by Foerster and Wattenhofer~\cite[Section~V, Theorem~3]{FoersterW16}.
A maximum-size first batch can still lengthen the eventual
schedule~\cite[Section~III-C]{FoersterLMS18}.
In our strong-safety model, this
objective is to maximize $|S|$ subject to acyclicity of $G_1(S)$.
Two rounds require one set $S$ making \emph{both} $G_1(S)$ and $G_2(S)$ acyclic.

The dependency-forest approach originates with Mahajan and Wattenhofer
\cite{MahajanW13}; Foerster, Mahajan, and Wattenhofer
\cite[Section~V]{FoersterMW16} study maximal safe batches and dynamic
dependency forests. A batching schedule is minimal if no individual later
update can be added to an earlier batch; this property does not imply a
minimum number of batches.

\begin{figure}[htbp]
\centering
\begin{tikzpicture}[x=1cm,y=1cm,
  batch/.style={draw=inkgrey,rounded corners=2pt,minimum height=8mm,
    text width=2.8cm,align=center,font=\small,inner sep=4pt},
  nextbatch/.style={batch,text width=3.5cm},
  flow/.style={-{Stealth[length=1.8mm]},line width=.7pt,inkgrey}]
  \node[figheading] at (2.6,.35) {Old and new successors};
  \node[anchor=north,font=\small] at (2.6,.05) {
    \begin{tabular}{@{}lcccc@{}}
      \toprule
      Router & $1$ & $2$ & $3$ & $4$\\
      \midrule
      \textcolor{oldblue}{Old} & $d$ & $d$ & $1$ & $2$\\
      \textcolor{neworange}{New} & $3$ & $3$ & $4$ & $d$\\
      \bottomrule
    \end{tabular}};
  \node[figheading] at (10.3,.35) {Why the bad remainder is unsafe};
  \node[routevertex] (one) at (8.7,-.65) {$1$};
  \node[routevertex] (three) at (11.9,-.65) {$3$};
  \draw[newedge] (one) to[bend left=22]
    node[above,figlabel] {new} (three);
  \draw[oldedge] (three) to[bend left=22]
    node[below,figlabel] {old} (one);
  \node[figlabel] at (8.7,-1.55) {already switched};
  \node[figlabel] at (11.9,-1.55) {still waiting};

  \node[figheading] at (1.65,-2.5) {First round};
  \node[figheading] at (6.1,-2.5) {All remaining updates};
  \node[figheading] at (11.55,-2.5) {Best completion};

  \node[batch,fill=earlyfill] (badfirst) at (1.65,-3.35)
    {$\{2,4\}$: safe};
  \node[nextbatch,fill=latefill,draw=red!65!black] (badrest) at (6.1,-3.35)
    {$\{1,3\}$: unsafe together};
  \node[align=center,font=\small,text width=3.8cm] (badend) at (11.55,-3.35)
    {\textbf{3 rounds total}\\finish with $\{3\}$, then $\{1\}$};
  \draw[flow] (badfirst)--(badrest);
  \draw[flow] (badrest)--(badend);

  \node[batch,fill=earlyfill] (goodfirst) at (1.65,-4.65)
    {$\{3,4\}$: safe};
  \node[nextbatch,fill=latefill] (goodrest) at (6.1,-4.65)
    {$\{1,2\}$: safe together};
  \node[align=center,font=\small,text width=3.8cm] (goodend) at (11.55,-4.65)
    {\textbf{2 rounds total}\\all updates finished};
  \draw[flow] (goodfirst)--(goodrest);
  \draw[flow] (goodrest)--(goodend);

  \node[figlabel] at (6.9,-5.55)
    {Both first rounds are largest possible: two routers.};
\end{tikzpicture}
\caption{Two maximum-size safe first batches can yield different schedule
lengths, as observed in~\cite[Section~III-C]{FoersterLMS18}.
This self-contained example admits the shown mixed cycle during round~2
after $\{2,4\}$.}
\label{r:first-choice}
\end{figure}
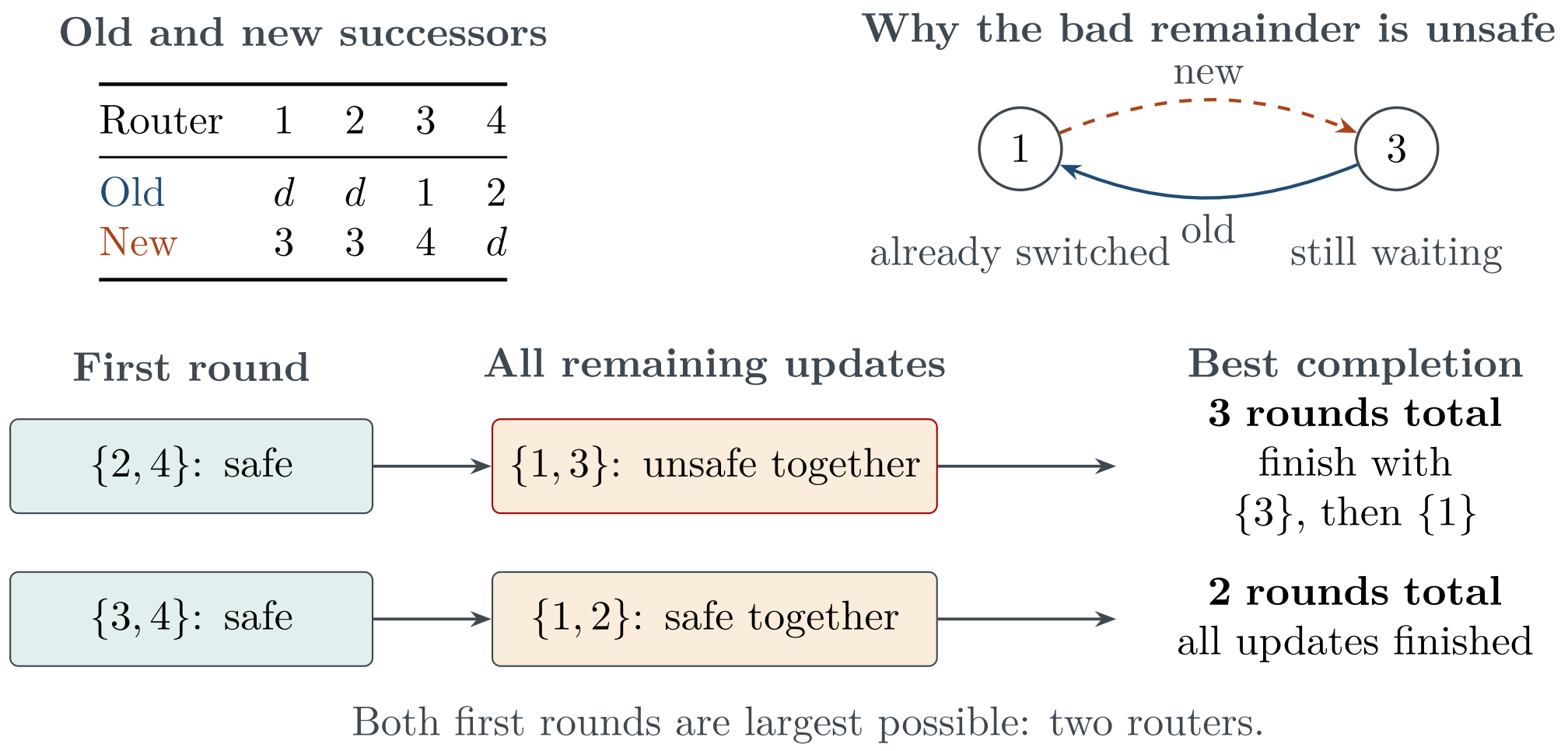

In Figure~\ref{r:first-choice}, the maximum safe first batches are exactly
$\{2,4\}$ and $\{3,4\}$. Router $1$ cannot belong to a safe first batch:
if it switches alone, $1\to3\to1$ is a cycle. A first batch containing both
$2$ and $3$ admits the cycle $2\to3\to4\to2$ while $4$ still uses its old
rule. After $\{2,4\}$, updating the remainder $\{1,3\}$
together is unsafe: if $1$ finishes first, $1\to3\to1$ is a cycle.
The completion $\{3\},\{1\}$ needs two further rounds. In contrast,
$\{3,4\},\{1,2\}$ is a safe two-round schedule. Thus the size of the first
batch does not determine the length of the remaining schedule.

\paragraph{Local verification and distributed updates.}
Foerster, Luedi, Seidel, and Wattenhofer~\cite[Section~5]{FoersterLSW18}
consider prescribed parent replacements between routing trees. Depth
labels let routers verify and perform loop-free updates
locally. Foerster and Schmid~\cite[Section~V]{FoersterS19} extend this
approach to successive tree migrations using version and depth labels,
without packet tagging. These protocols provide local certification and
execution of safe tree updates. Our two-round problem concerns a single
prescribed migration with globally safe batches. Reducing controller
communication can shorten execution time even when replacements remain
sequentially dependent
\cite[Section~VI]{FoersterS19}.
P4Update develops local verification and dependency notifications into
a P4 data-plane implementation for consistent updates~\cite{ZhouHKBF21}.

\paragraph{Multicast parent migration.}
Shifter and Nous study parent replacements in multicast
routing~\cite{GaoMGWJKC24,WuGCCKC18Shifter}. There is a useful graph
analogy, but it depends on the exact safety requirement. Reverse every
arc of our two spanning in-arborescences. The resulting graphs are
out-arborescences with a common root; each nonroot vertex now replaces
its parent rather than its successor. For any completed set $T$,
\[
 F(T)\text{ is acyclic}
 \quad\Longleftrightarrow\quad
 d\text{ reaches every vertex in }F(T)^{\mathrm{rev}}.
\]
Indeed, every nonroot vertex has exactly one successor in $F(T)$, so
a successor walk either reaches $d$ or enters a cycle. The equivalence
holds for every partial completion and therefore preserves round counts
for the \emph{abstract} all-target, fixed-vertex-set parent-replacement
problem. It does not by itself identify that abstract problem with every
operation and additional drop- or duplicate-freedom requirement of a
particular multicast protocol.

\paragraph{Several destinations and policies.}
Router-wide migration may change routes for several destinations at once
\cite{VanbeverVPPB12}; overlapping policies can share forwarding rules
\cite{YuBZFL18,YuFZB17}. The scheduling choices then interact across logical
routes. A hardness reduction that uses these interactions need not survive
the restriction to one destination and one prescribed successor replacement
per router. Dudycz, Ludwig, and Schmid~\cite{DudyczLS16} show that
minimizing switch interactions is NP-hard already for two routing policies,
even when each policy can individually be updated in two rounds.
The number of policies and the number of asynchronous rounds are
different parameters.

For multiple unsplittable flows, Henzinger et al.\ study the tradeoff
between update rounds and temporary link-capacity
augmentation~\cite{HenzingerPPS22}. Their model includes congestion
constraints, which are absent here.

\paragraph{Versioned updates.}
The versioned mechanism of Reitblatt et al.\ first installs new rules
alongside the old ones and then switches the version selected by incoming
packets~\cite[Sections~2 and~5]{ReitblattFFMW12}. Packet tags select the
appropriate rules and provide per-packet consistency, while routers may
temporarily retain both versions. The two protocol phases thus permit
operations beyond the prescribed replacements studied here.

\paragraph{Limitations.}
The reduction does not bound router degrees: many branches meet at $g$
or $d$, and a blocker can receive edges from many occurrences of the
same literal. Theorem~\ref{thm:main} therefore does not establish
bounded-degree hardness. The result is also restricted to strong loop
freedom and to the prescribed-rule model; it does not classify
chosen-ingress relaxed loop freedom for trees.
\paragraph{AI assistance.}
I used ChatGPT (Astra~6.2) for research, drafting, figures, and work on the
companion Lean, Coq, and Isabelle developments, including their encoding
and complexity extensions. Claude (Fable~5.2) provided referee-style
reviews of the direct 3-SAT argument and the proof artifact.

{\small
\interlinepenalty=10000
\bibliographystyle{arxivdoi}
\bibliography{references}
}

\clearpage
\appendix
\section{Explicit ranks for the two-round construction}\label{app:ranks}

Fix a satisfying assignment and the first-round set $S$ from
Table~\ref{p:assignment-schedule}, including $g$. Truth values refer to
this assignment; all closers and connectors are late. We give ranks,
zero at $d$, that strictly decrease on every edge of $G_1(S)$ and
$G_2(S)$, proving the acyclicity claimed in Section~\ref{p:sufficient}.
The decreasing-rank proofs in \texttt{lean/arxiv/ArxivSAT/Sufficiency.lean}
use different numerical offsets.

\paragraph{First round.}
Let $t(x)$ be the number of old spine edges on the path from a spine
vertex $x$ to $d$. Thus $t(C_m)=1$ and $t(g)=4m+1$. Define
\[
\begin{array}{@{}ll@{\qquad}ll@{}}
\toprule
\text{Router }x & \rho_1(x) & \text{Router }x & \rho_1(x)\\
\midrule
 d &0 & Z_\ell\text{ (any) }&1\\
 B_\ell\text{ (true) }&2 & x\text{ on spine},\ x\ne d&2+t(x)\\
 B_\ell\text{ (false) }&4m+4 & &\\
\bottomrule
\end{array}
\]
Every old spine edge decreases $t$, with $C_m\to d$ going from rank
$3$ to $0$. Old branches $B_\ell\to Z_\ell\to d$ also decrease.
The new edges in $G_1(S)$ go from rank $4m+3$ to $0$ at $g$, from
at least $4$ to $2$ at true-literal choices, and from $4m+4$ to
$4m+3$ at false-literal blockers. Thus every edge decreases $\rho_1$.

\Needspace{16\baselineskip}
\paragraph{Second round.}
For each clause $i$, choose one true literal position
$j_i^*\in\{1,2,3\}$. Put
$\operatorname{pos}(R_{i,j})=j$ and $\operatorname{pos}(C_i)=4$.
Write $[z]_4\in\{0,1,2,3\}$ for the residue of an integer modulo four.
Define the following rank:
\[
\begin{array}{@{}ll@{\qquad}ll@{}}
\toprule
\text{Router }x & \rho_2(x) & \text{Router }x & \rho_2(x)\\
\midrule
 d&0 & g&1\\
 B_\ell\text{ (false) }&2& Z_\ell\text{ (true) }&3\\
 B_\ell\text{ (true) }&4& Z_\ell\text{ (false) }&5\\
\multicolumn{4}{@{}l@{}}{\displaystyle
 x\in\{R_{i,1},R_{i,2},R_{i,3},C_i\}:\quad
 \rho_2(x)=5+4(m-i)+[j_i^*-\operatorname{pos}(x)]_4.}\\
\bottomrule
\end{array}
\]
The new edges of $g$, blockers and connectors decrease these ranks:
false-literal blockers go to $g$; true-literal blockers go to $g$;
a true-literal connector goes to the opposite false blocker, and a
false-literal connector goes to the opposite true blocker. The old
edges retained by late blockers and connectors decrease rank as well.
Every choice's new edge enters a blocker of rank at most four, while
its own rank is at least five.

Within clause $i$, an old choice edge advances from position $j$ to
$j+1$. Its rank drops by one unless $j=j_i^*$; at that position the
choice is early and its old edge is absent from $G_2(S)$. The closing
new edge $C_i\to R_{i,1}$ also decreases rank by one. Finally, each
closer's retained old edge goes to the next clause, whose base rank is
four lower, or to $d$ for the last clause. Therefore every edge of
$G_2(S)$ decreases $\rho_2$. In both rounds, all ranks belong to
$\{0,\ldots,4m+4\}$.

The argument does not require distinct literals within a clause:
choice routers are indexed by positions and share blockers where
appropriate. Neither tautological clauses nor repeated occurrences
change the rank inequalities.

\clearpage
\section{Automated Proofs}\label{p:formalization}

\paragraph{Routing and safety.}
Lean~4.35.0, Coq~8.20.1 and Isabelle2026 prove the direct 3-SAT
reduction for every partial completion of each round. For nonempty formulas
they prove the endpoint trees, exact size, four-step bound on new routes,
and optimum two on satisfiable instances and three on unsatisfiable ones.
Lean also checks the two rank tables in Appendix~\ref{app:ranks}, including
the $4m+4$ bound. Isabelle gives a second sufficiency proof by tracing
forwarding walks in the two round unions.
Fresh Isabelle builds replayed the graph proof, the encoding and checker,
the equivalence of bit languages, and the reconstructed automation.
The audits found no oracle or skipped-proof dependencies.

\paragraph{Encodings and complexity.}
Isabelle and Coq read two counted columns of unary integers for the old
and new successors. A supplied input adds a third column with a proposed
round label for each router. The destination has index zero. Isabelle identifies
its ordinary and supplied bit languages with predicates given by shared
decoding and certificate equations. Coq~8.16.1 proves that decoding
preserves and reflects the corresponding predicates on typed tables and
schedules. It proves polynomial-time reductions in both directions and hence
NP-completeness for both bit languages. The supplied decision language
requires the given three-round schedule to be safe \emph{and} a safe
two-round schedule to exist. Coq counts beta steps on Scott-encoded inputs
in the weak call-by-value calculus~L of the Complexity
library~\cite{GaeherK21}, whose Cook--Levin theorem supplies 3-SAT NP-completeness.\footnote{\raggedright
\url{https://github.com/uds-psl/coq-library-complexity}, commit
\nolinkurl{14b5f413d2fb7adecde79c5451b483f9a1af59a8}.}
Isabelle proves a quadratic parser and certificate-checker bound under its
own explicit charges.

\paragraph{Returning a schedule.}
All three graph developments verify the $3/2$ algorithm in
Section~\ref{r:models}. Coq also checks a four-router instance on which it
returns three rounds and the optimum is two. The Coq~8.16.1 theorem
\nolinkurl{encoded_three_halves_end_to_end} starts with the supplied input
bits. For a string of $n$ bits, decoding, testing the union and returning the
labels together take $O((n+1)^4)$ beta steps in L, including on malformed
inputs. If the endpoints are valid routing trees and the supplied
three-round schedule is safe, the output consists of a count $r$ and one
zero-based round label per non-destination router. Every partial completion
is safe and $2r\leq 3\,\mathrm{OPT}$. The destination alone gives zero
rounds and an empty list. Otherwise the algorithm returns one round if the union is acyclic.
If it contains a cycle, the algorithm returns the supplied three rounds
with the destination label removed, including whenever $\mathrm{OPT}=3$. Unequal successor-column lengths, trailing
bits or any other decoding error give zero and an empty list. For a parsed
input the safety and ratio conclusions require the stated hypotheses.

\paragraph{The strict threshold.}
The Coq~8.16.1 theorem \nolinkurl{wire_sharp_three_halves} verifies
Corollary~\ref{cor:approx} in L on the same supplied bit strings.
A polynomial-time solver that returns a safe schedule on every valid supplied
input with a fixed rational guarantee $a/b$, where $b>0$ and $2a<3b$,
must return at most two rounds on satisfiable reduction
instances and at least three on unsatisfiable ones. Testing its count would
decide 3-SAT in polynomial time and put every NP predicate on an efficiently
encodable type in P. At equality, three rounds are possible in both cases.
Lean reports \texttt{propext}, \texttt{Classical.choice} and
\texttt{Quot.sound}. The Coq audits identify uses of excluded middle and
definite description.

\paragraph{Companion artifact.}
The proof sources, recursive Coq kernel replay records and assumption audits
are available in the public companion repository at\\*
\url{https://github.com/ktfoerster/two-rounds-are-hard}.

\end{document}